\documentclass{article}[12pt]
\usepackage[utf8]{inputenc}
\usepackage{graphicx}
\usepackage{url}
\usepackage{hyperref}
\usepackage{float}
\usepackage{placeins}
\usepackage{authblk}
\usepackage{xcolor}
\usepackage{geometry}
\usepackage{algorithm} 
\usepackage{algpseudocode} 

\title{Intermunicipal Travel Networks of Mexico (2020-2021)}

\author{Oscar Fontanelli$^{1}$, Plinio Guzmán$^{2}$, Amilcar Meneses$^{3}$, Alfredo Hernández$^{4}$, Marisol Flores-Garrido$^{5}$, Maribel Hernández-Rosales$^{1\ast}$, Guillermo de Anda-Jáuregui$^{6,7,8,\ast}$}
\date{\small
    $^1$ CINVESTAV, Irapuato, Mexico\\[0.5ex]%
    $^6$ Computational Genomics Division, National Institute of Genomic Medicine, Mexico City, Mexico\\%
    $^7$ Cátedras Conacyt Para Jóvenes Investigadores, National Council on Science and Technology, Mexico City, Mexico\\[0.5ex]%
    $^8$ Centro de Ciencias de la Complejidad (C3), Universidad Nacional Autónoma de México , Mexico City, Mexico\\[0.5ex]%
    $^2$ Astronomer LTD\\[0.5ex]%
    $^3$ CINVESTAV CDMX, Mexico\\[0.5ex]%
    $^4$ Centro de Ciencias Genómicas, UNAM\\[0.5ex]%
    $^5$ ENES Morelia, UNAM\\[1ex]%
    $^\ast$ \url{gdeanda@inmegen.edu.mx}\\
    $^\ast$ \url{maribel.hr@cinvestav.mx}\\
    }

\begin{document}

\maketitle

\begin{abstract}
We present a collection of networks that describe the travel patterns between municipalities in Mexico between 2020 and 2021. Using anonymized mobile device geo-location data we constructed directed, weighted networks representing the (normalized) volume of travels between municipalities. We analysed changes in global (graph total weight sum), local (centrality measures), and mesoscale (community structure) network features. We observe that changes in these features are associated with factors such as Covid-19 restrictions and population size. In general, events in early 2020 (when initial Covid-19 restrictions were implemented) induced more intense changes in network features, whereas later events had a less notable impact in network features. We believe these networks will be useful for researchers and decision makers in the areas of transportation, infrastructure planning, epidemic control and network science at large. 
\end{abstract}

\section*{Introduction}
Intermunicipal mobility is a kind of medium and large scale mobility within a country where millions of individuals daily travel from one county or municipality to another, either going from home to work, shopping, accessing  public services, cargo loading, vacation, etc. These movements and travels generate complex structures and dynamics of socio-economic interactions between different areas both at regional and national levels.

Given the nature of these mobility systems, complex networks have been widely adopted to model this commuting phenomena \cite{caschili2013accessibility}\cite{de2010modeling}\cite{lenormand2012universal}\cite{lotero2016rich}\cite{ramasco2009using}. Characterizing and understanding the properties of these mobility networks is crucial for decision-making, urban planning, traffic engineering and, as has become clear with the Covid-19 pandemic, designing, implementing and evaluating mobility restrictions and lockdowns in order to contain or control the epidemic spread  \cite{bai2020mapping}\cite{fajgelbaum2021optimal}\cite{melo2021heterogeneous}\cite{pullano2020population}\cite{seto2020commuting}\cite{sun2020quantifying}\cite{yilmazkuday2020covid}. Therefore the need for public mobility network datasets, aiming to researchers and decision-makers in the areas of mobility, urban planning, epidemic control, etc. 

Traditionally these networks had been obtained from mobility surveys. However, the emergence in recent years of cell phone and GPS data have facilitated the acquisitions of accurate and large sets of human mobility data, thus allowing the construction and characterization of large and detailed human mobility networks. 

Here we introduce a new public dataset of daily intermunicipal origin-destination networks in Mexico for 2020 and 2021, which were directly constructed from large datasets of geolocation data. With this dataset we hope to contribute to research and decision-making communities from diverse interests, from pure network theory to those studying human mobility, urban planning, national scale social and economic relations, epidemic control, etc. As far as we know, this is the first public dataset of its kind for a country in Latin America.

There are other works that have built origin-destination networks aiming to describe mobility patterns, measure the volume of public transportation and to plan public transportation, among other things. These mobility patters include large-scale and long-range commuting patterns~\cite{hadachi2020unveiling}\cite{riascos2020networks}; spatio-temporal patterns for different socioeconomic strata~\cite{lotero2016rich}; patterns in bike-sharing systems~\cite{loaiza2019human}, etc. Origin-destination matrices have also been used to measure the volume of use of public transport~\cite{alsger2015use}\cite{wang2019origin}\cite{munizaga2012estimation} and  public transport planning~\cite{nasiboglu2012origin}\cite{ait2021value}. 

There are several approaches for modeling and generating origin-destination matrices. These approaches include gravity models~\cite{tolouei2017Origin}\cite{simini2012universal}\cite{ekowicaksono2016estimating},Bayesian model~\cite{perrakis2012bayesian}\cite{pitombeira2020dynamic}\cite{wang2013sensor}, linear assignment matrix approximation~\cite{toledo2012estimation}, Principal Components Analysis~\cite{djukic2012efficient} and gradient approximation method~\cite{frederix2011new}, among others. 

Origin-destination networks have been elaborated from different types of data. Most of these works use data from the different public transport systems or from road side interview~\cite{tolouei2017Origin}. In works such as~\cite{alsger2015use} and \cite{munizaga2012estimation} authors use smart card fare data to estimate origin-destination networks. In~\cite{lotero2016rich} the data considered for these matrices are those of the bike-shared system. In~\cite{wang2019origin} authors conduct experiments on two datasets generated by ride-hailing applications. In~\cite{toledo2012estimation}, authors estimate the matrices using data from traffic counts on the network links. 

Recently, other works have attempted to utilize social network data from Twitter or Facebook~\cite{osorio2019social}\cite{pourebrahim2019trip}\cite{bonnel2018origin}\cite{bonaccorsi2020economic}\cite{edsberg2022understanding} \cite{galeazzi2021human} or another data set from mobile phone data such ACAPS dataset or SafeGraph~\cite{schlosser2020covid}\cite{chang2021mobility}\cite{deng2021network}. In particular, Edsberg et al.~\cite{edsberg2022understanding} show that origin-destination networks based on data from cell phones and social networks present high-quality results. 

This article is organized as follows: Results presents and describes the collection of mobility networks and it shows  analysis of changes in global (sum of weights), local (centrality measures), and mesoscale (community structure) network features. In Discussion we argue how events in early 2020 (when initial Covid-19 restrictions were implemented) induced more intense changes in network features, whereas later events had a less notable impact in network features. In Methods we describe the methodology we utilized to collect data and show the algorithm for network construction. 
\section{Results}

\subsection*{A travel network dataset}
We release a  public dataset of 731 intermunicipal origin-destination networks in Mexico. These networks were constructed from a large and anonymized mobile location dataset (see \cite{de2020contact} for more information about this dataset). Each network is the intermunicipal origin-destination network in Mexico for each day during the 2020-2021 period. Nodes represent municipalities (third level administrative division) or official metropolitan zones, see Methods section below. These are weighted and directed networks, where the weight of edge $a_{ij}$ is equal to the total number of observed travels from node $i$ to node $j$ normalized by the different number of mobile devices we recorded on that day. The data set with these 731 networks is freely available in a OSF repository \url{http://dx.doi.org/10.17605/OSF.IO/42XQZ}. 

\begin{figure}[!t]
    \centering
    \includegraphics[width=\textwidth]{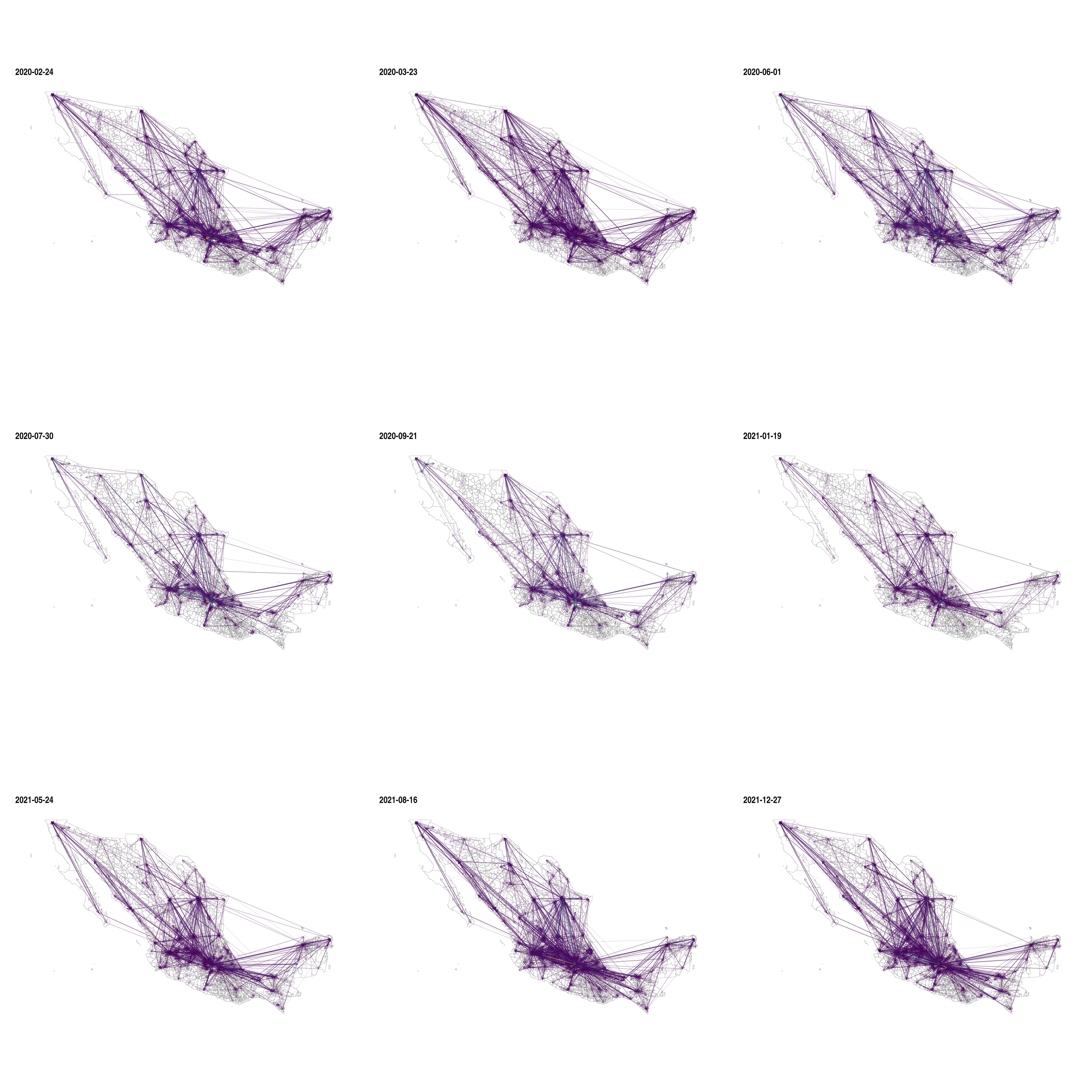}
    \caption{Network visualizations for selected days.}
    \label{fig:mapa_nw}
\end{figure}

For analysis and visualization purposes we chose nine representative dates capturing different important events during the evolution of the pandemic in Mexico; these are shown in Table \ref{table-dates}. As a first visualization, we show in Fig. \ref{fig:mapa_nw} mobility networks over the Mexico map for this set of dates, marking only 1$\%$ of edges with highest weight.

\FloatBarrier
\begin{table}[!t]
\centering
\begin{tabular}{|l|l|}
\hline
Date       & Event                                                                   \\ \hline
2020-02-24 & First reported Covid-19 case.                                           \\
2020-03-23 & Beginning of official lockdowns (National Program of Social Distance). \\
2020-06-01 & Beginning of Epidemiological Stoplight Program.                         \\
2020-07-30 & First national peak of daily contagions.                                \\
2020-09-21 & Local minimum of daily contagions (between first and second wave).                                             \\
2021-01-19 & Second national peak of daily contagions.                               \\
2021-05-24 & Local minimum of daily contagions (between second and third wave).                                             \\
2021-08-16 & Third national peak of daily contagions                                 \\
2021-12-27 & Local minimum of daily contagions (just before fourth wave)                    \\ \hline
\end{tabular}
\caption{Set of dates for the analysis. We chose these dates to be either Monday or the Monday closest to the referred event. All of these make reference to events in Mexico.}
\label{table-dates}
\end{table}

\subsection*{Changes in graph total weight sums were observed throughout the first two years of the pandemic}
As a measure to quantify the total observed movement in each network of the collection, we consider the total sum of weights in the network $S_{G} = \sum w_i$, where $i$ runs over all nodes of the network. In our context, a higher value of $S_{G}$ can be understood as higher mobility between municipalities in the country, which, in turn, is associated with people's decisions to move outside their locality. 

\begin{figure}[!t]
    \centering
    \includegraphics[width=\textwidth]{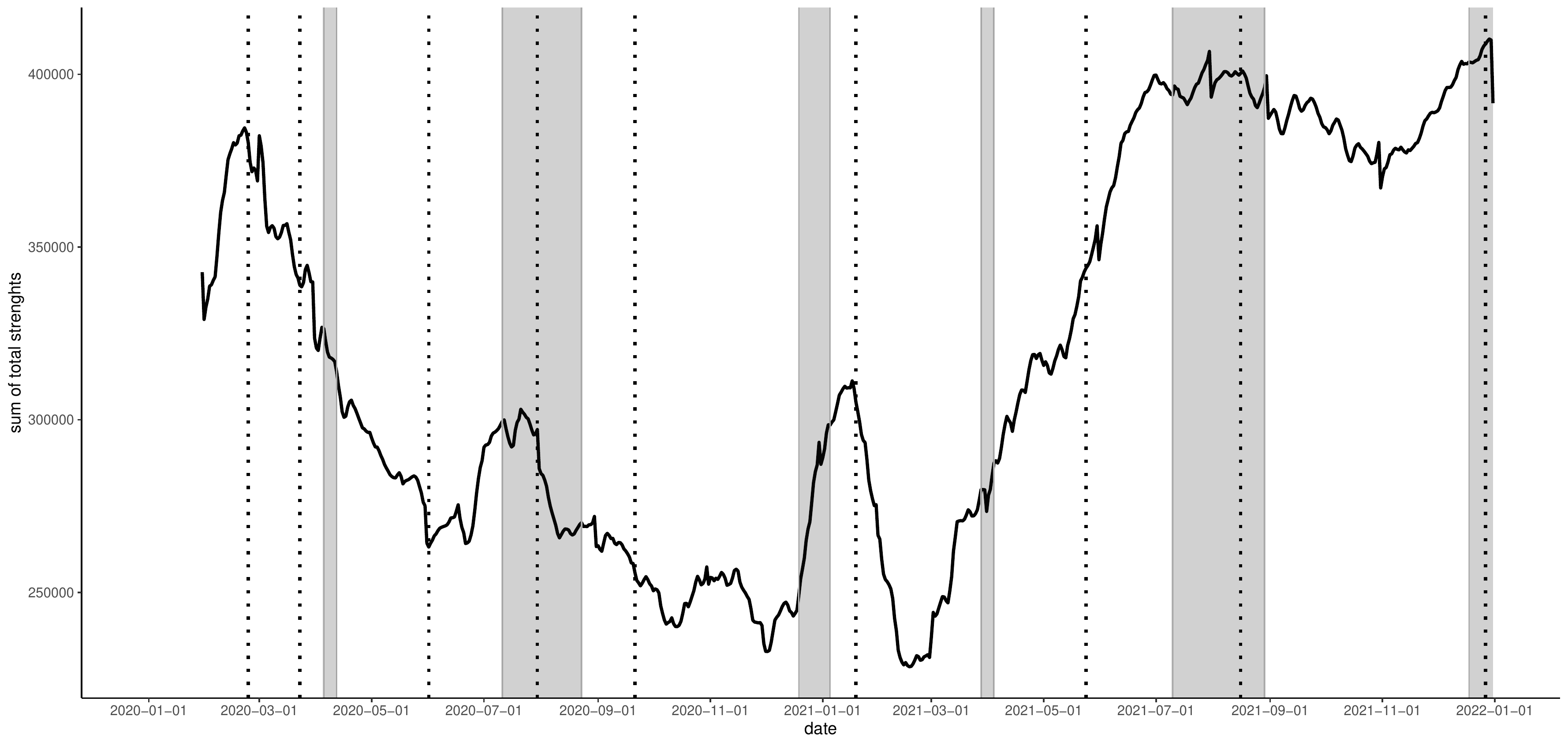}
    \caption{Time series of smoothed edge weight sum (weighted density). Vertical dotted lines correspond to dates shown on Table \ref{table-dates}. Shaded areas are official school holiday periods.}
    \label{fig:ts}
\end{figure}

Fig. \ref{fig:ts} shows the time series related to the $S_{G}$ parameter, also indicating the set of dates described in Table \ref{table-dates} and official school vacation periods (summer breaks, winter breaks and Easter holidays, shaded in gray).  It can be observed that the decay in mobility that happens in January 2020, probably due to being the post-holiday season, is prolonged after the start of the lockdown, reaching a local minimum point shortly before the beginning of the summer holiday season. Following summer break mobility continues to decay, reaching its lowest point again shortly before the beginning of winter break, when there was a pronounced mobility rise. For the first half of 2021 we observe a sustained rise in mobility until July 2020, when it reaches a relatively high plateau. 

Since the networks exhibit a different mobility pattern each day, centrality measures associated with each node (locality) also change. In this section we explore the variability of centrality measures over the period we studied. We choose three different centrality measures: (a) degree centrality, (b) node strength, (c) betweenness centrality. Variability is measured through the coefficient of variation (cv). 

\begin{figure}[htp]
    \centering
    \includegraphics[height=10cm]{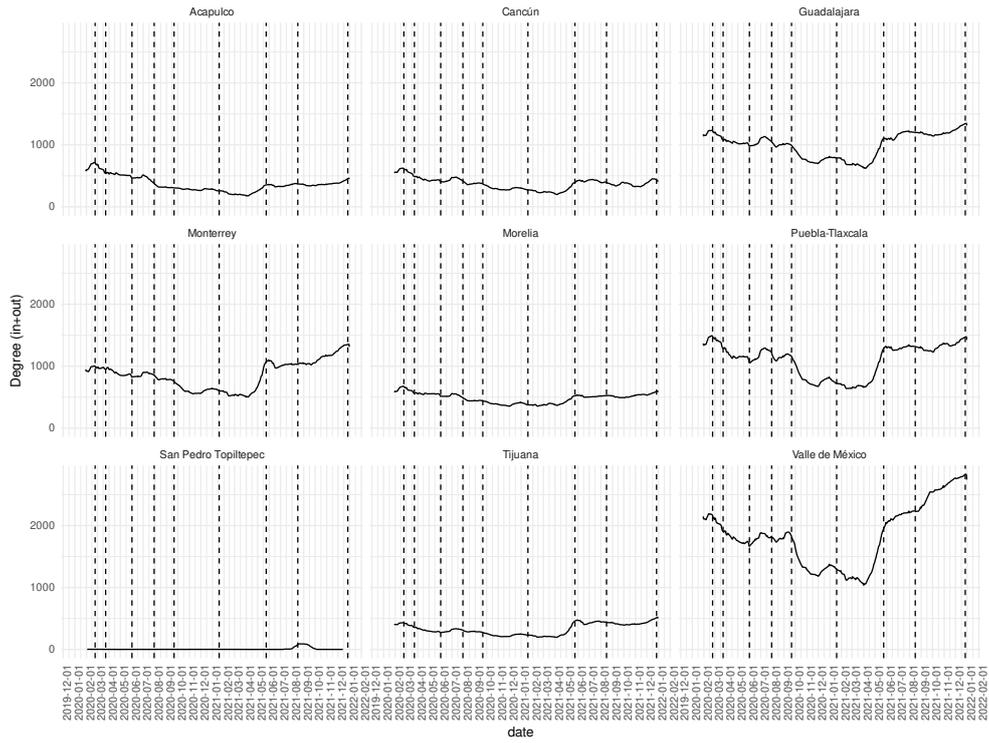}
    \caption{Time series of total degree for nine representative nodes.}
    \label{fig:grado}
\end{figure}

\begin{figure}[htp]
    \centering
    \includegraphics[height=10cm]{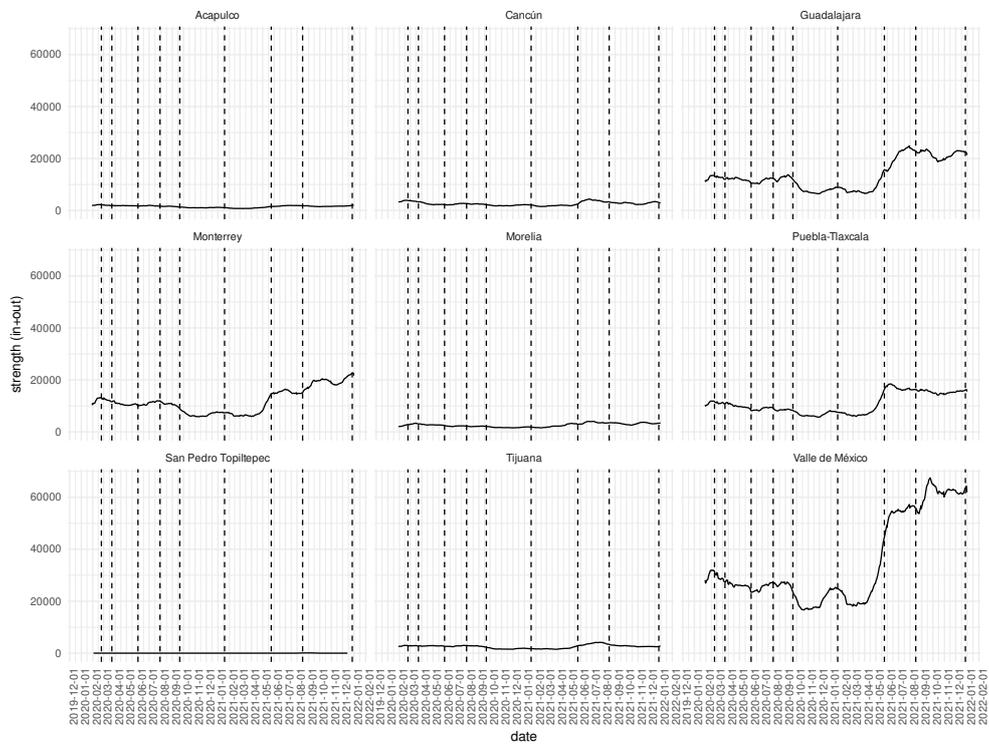}
    \caption{Time series of total strength for nine representative nodes.}
    \label{fig:my_fuerza}
\end{figure}

\begin{figure}[htp]
    \centering
    \includegraphics[height=10cm]{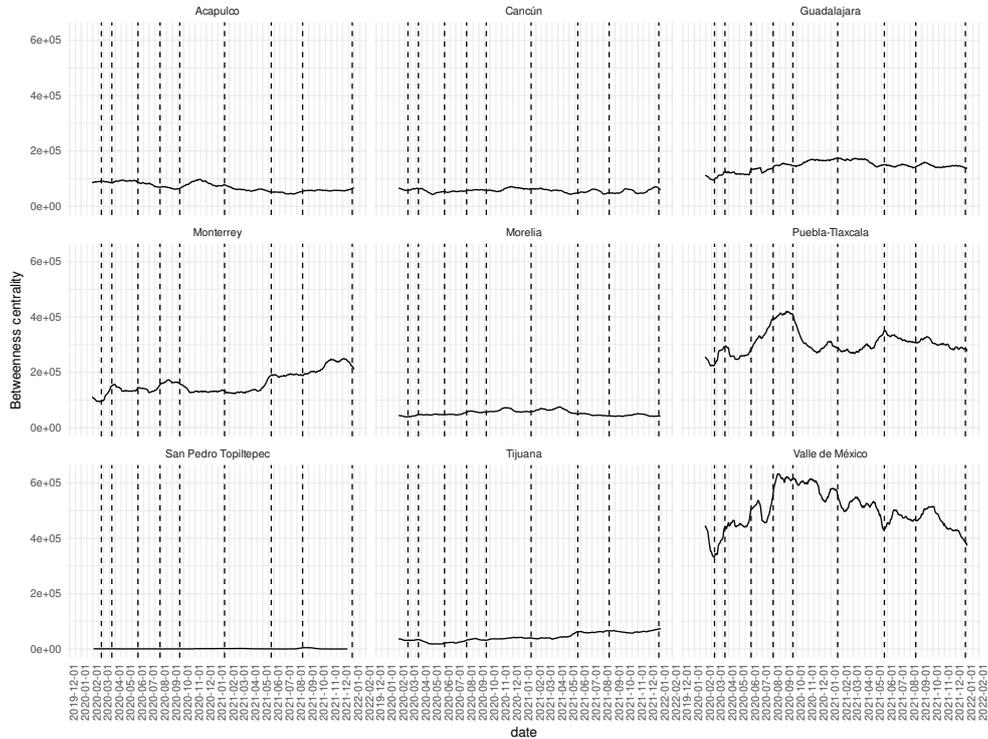}
    \caption{Time series of betweenness centrality for nine representative nodes.}
    \label{fig:my_betweenness}
\end{figure}

Figures \ref{fig:grado}-\ref{fig:my_betweenness} show the variation of centrality for nine different nodes (municipalities or metropolitan zones). These nodes were chosen to illustrate different behaviours across the studied time period. 

The regions of Valle de México, Guadalajara, Monterrey and Puebla-Tlaxcala correspond to densely populated regions. In these areas, the coefficients of variation for the three different centrality measures are: (30.0, 53.1, 9.7) in Valle de México, (24.3, 50.3, 17.8) in Guadalajara, (31.3, 49.2, 25.9) in Monterrey, and (29.0, 44.0, 34.7) in Puebla-Tlaxcala. Morelia, a city in west-central Mexico, is included because it maintains a highly stable mobility pattern with coefficients of variation 24.0, 40.3 and 16.0 for the considered centrality measures. The municipality of San Pedro Topiltepec is an example of a locality with high degree-centrality variability. The coefficients of variation for this location are 320.4, 410.1 and 285.8. Tijuana is an important border city and changes in centrality are expected whenever there are changes in the regulations on the border with the USA. The coefficients of variation are 34.9, 41.5 and 89.6. Finally, Acapulco and Cancun are cities with a beach that represents a popular tourist destination. The coefficients of variation for these cities are (41.4, 41.1, 60.2) in Acapulco, and (33.3, 41.8, 62.1) in Cancún. 


\FloatBarrier

\begin{figure}[htp]
    \centering
    \includegraphics[height=10cm]{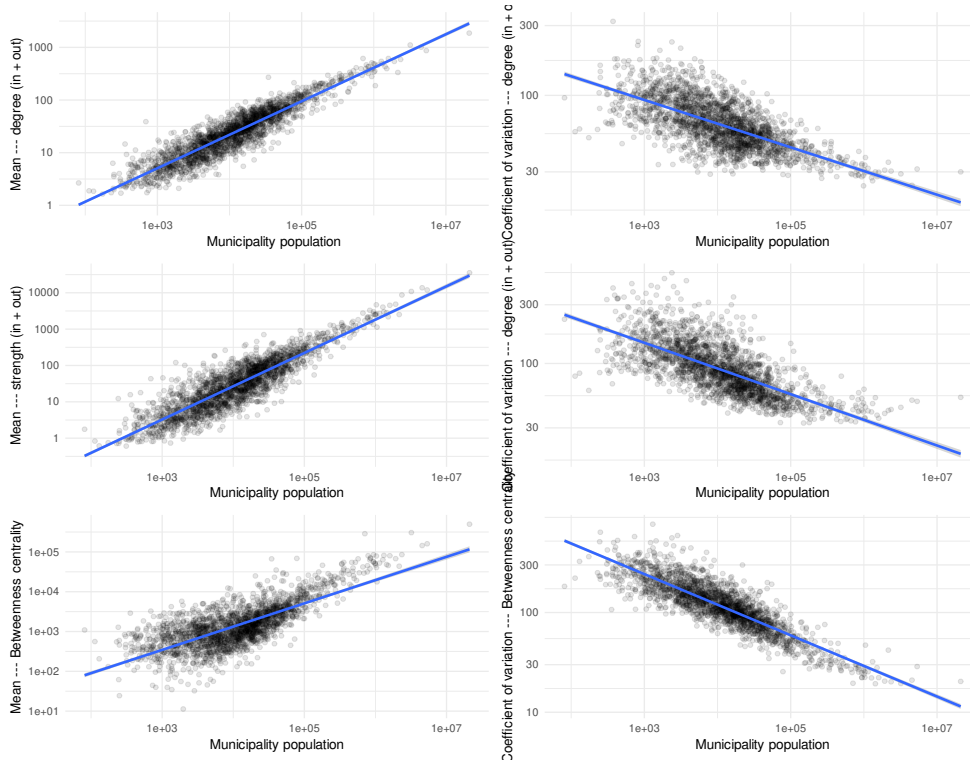}
    \caption{left: log-log relationship between mean of daily centrality measures and population (degree, strength, and betweenness centrality); right: log-log relationship between coefficient of variation of daily centrality measures and population (degree, strength, and betweenness centrality)}
    \label{fig:centralities_vs_pop}
\end{figure}

In order to explore the variability observed in node centrality measures with the population of the area they represent, the relationship between the three centrality measures considered and the population reported in the 2020 national census was analyzed \url{https://www.inegi.org.mx/programas/ccpv/2020/default.html#Datos_abiertos}. Table \ref{table:linearFit} shows estimated parameters when fitting a linear regression model, while Fig. \ref{fig:centralities_vs_pop} shows the identified relationship on a logarithmic scale.

\begin{table}[tp]
\centering
\label{table:linearFit}
\begin{tabular}{rlllr}
  \hline
 & Node metric & Statistic & Model& R squared \\ 
  \hline
1 & strength\_in & mean & lineal & 0.94 \\ 
  2 & strength\_all & mean & lineal & 0.93 \\ 
  3 & strength\_out & mean & lineal & 0.93 \\ 
  4 & degree\_out & mean & loglog & 0.84 \\ 
  5 & degree\_all & mean & loglog & 0.84 \\ 
  6 & degree\_in & mean & loglog & 0.84 \\ 
  7 & betweenness\_undirected & mean & lineal & 0.74 \\ 
  8 & betweenness\_directed & mean & lineal & 0.69 \\ 
   \hline
9 & betweenness\_undirected & cv & loglog & 0.75 \\ 
  10 & betweenness\_directed & cv & loglog & 0.74 \\ 
  11 & strength\_out & cv & loglog & 0.50 \\ 
  12 & degree\_out & cv & loglog & 0.50 \\ 
  13 & strength\_in & cv & loglog & 0.50 \\ 
  14 & strength\_all & cv & loglog & 0.48 \\ 
  15 & degree\_in & cv & loglog & 0.48 \\ 
  16 & degree\_all & cv & loglog & 0.44 \\ 
   \hline
\end{tabular}
\caption{Results for linear regressions for node metrics (mean and coefficient of variation) against population. Some of them where in lineal scale and some other in log scale.}
\end{table}

\subsection*{Formation and evolution of communities in the network.}
Recalling that a community in a network is a set of nodes with a larger density of connections between them than external to the set, communities in these mobility networks correspond to groups of municipalities or with a high internal (within the group) mobility and a relative low external mobility (from the region to the outside or the other way around). Community detection algorithms in mobility or commuting networks have been widely utilized to detect or delimitate geographical functional regions \cite{drobne2020comparison}\cite{duranton2015delineating}\cite{he2020demarcating}\cite{hong2019hierarchical}\cite{mu2020regional}.

\begin{figure}[htp]
    \centering
    \includegraphics[height=8cm]{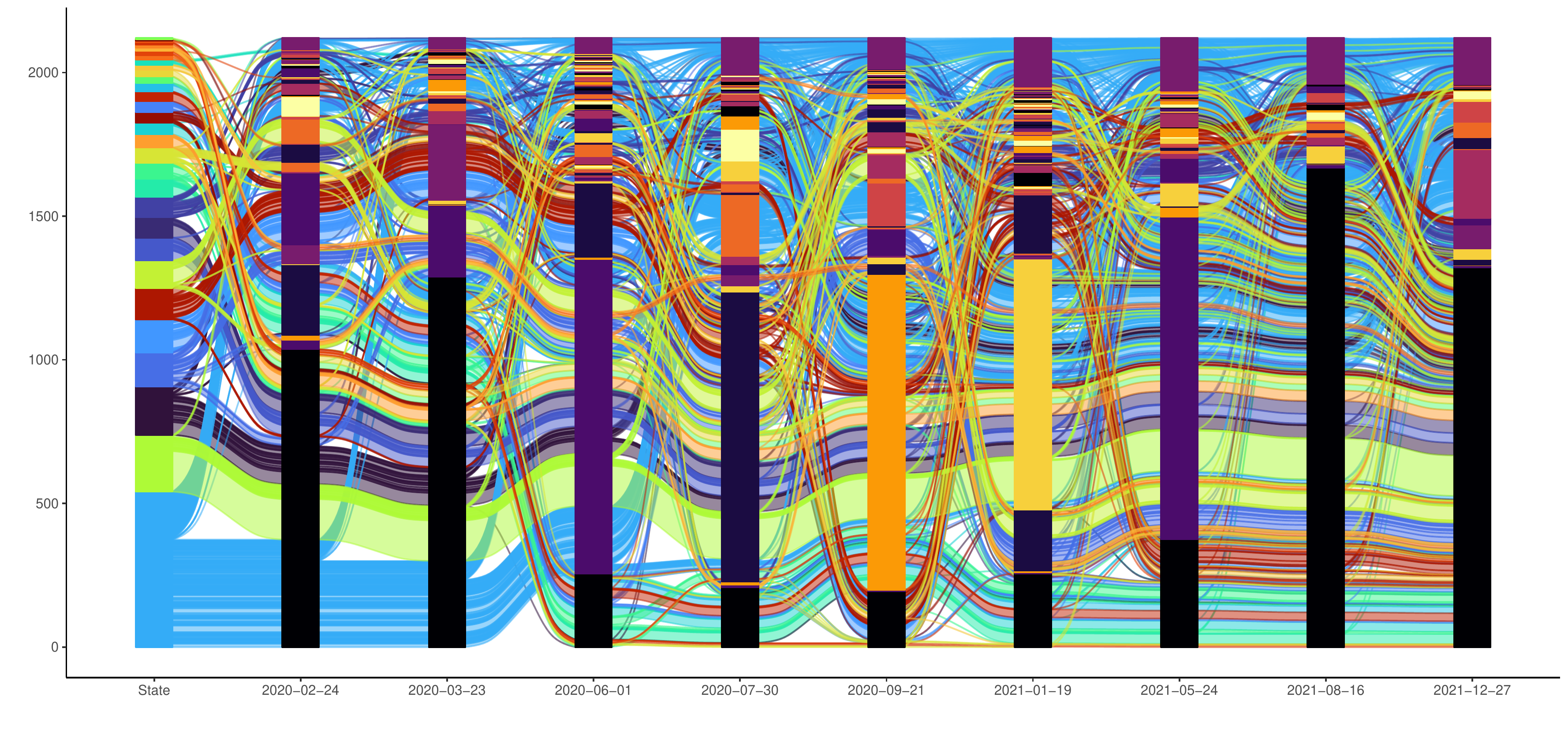}
    \caption{Alluvial plot for network communities for the set of nine relevant dates. Each line represents a node (municipality or metropolitan zone), they are coloured according to state and for each day they are grouped according to their community. For each day communities are shown in the vertical bars.}
    \label{fig:alluvial_national}
\end{figure}

We utilized the label-propagation algorithm to detect communities in our networks. Alluvial plot in Fig. \ref{fig:alluvial_national} shows in schematic way the time evolution of community structure on the networks. In this diagram each line represents a municipality in Mexico and they are grouped according to the network community to which they belong. We show this structure for the chosen set dates. Color of each line represents the state to which they belong.

\begin{figure}[!h]
    \centering
    \includegraphics[height=8cm]{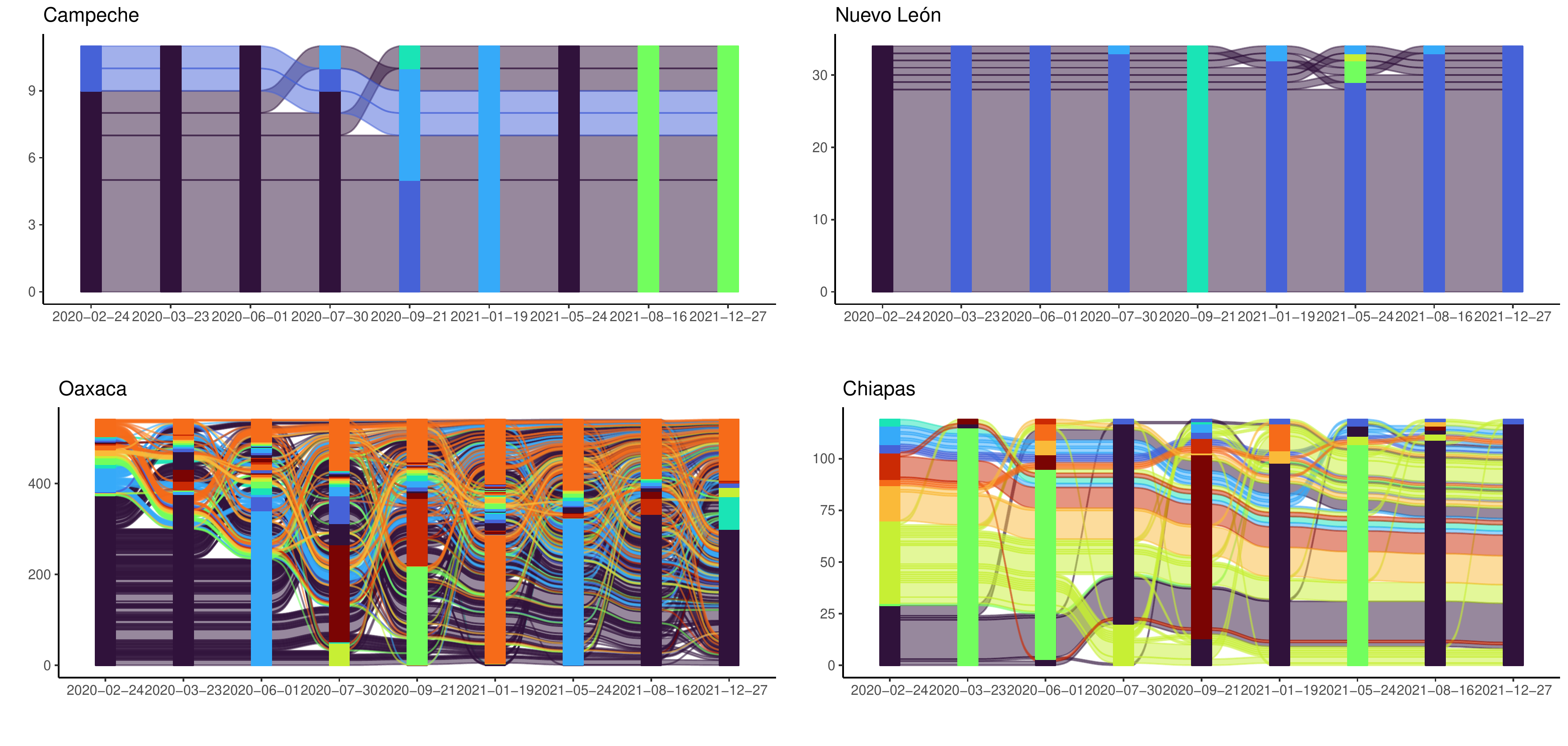}
    \caption{Alluvial plots for network community inside four states in Mexico. Each line represents a municipality or metropolitan zone, they are coloured according to their initial community and they are grouped according to the community to which they belong.}
    \label{fig:alluvial_states}
\end{figure}

\section*{Discussion}



Sum of total (in and out) weights over all edges of these mobility networks is a gross indicator of the total intermunicipality mobility that we observe on each day. We can see in Fig \ref{fig:ts} how there is a pre-pandemic peak at late February - early March 2020 (first confirmed case in Mexico was on February 24, 2020). This peak is followed by a persistent decay, whose rate does not seem to be affected by the beginning of lockdowns. In fact, this decay continues until June-July of 2020, when we observe a peak that somewhat coincides with official summer break.  After summer break we still see a downward trend until winter break and, agian, we see a peak that coincides with official vacation period. In late February 2021 we observe the lowest point of mobility and from here there is  sustained upward train, apparently unaffected by Easter holidays until it reaches a new maximum that is even higher than the pre-pandemic peak.  At this point mobility reaches a high plateau and, even though mobility still fluctuates, these fluctuations does not seem to be correlated neither with the pandemic third wave nor with vacation periods. From July 2021 total intermunicipal mobility stays at high levels, comparable with pre-pandemic levels.\\ 



Centrality measures evolve over time. How they change seem to be weakly associated to the evolution of the pandemic. We also observe that not all nodes exhibit the same behavior. While some nodes, such as Guadalajara, Puebla-Tlaxcala and Valle de Mexico show larger fluctuations in their degree centrality, other nodes such as San Pedro Topiltepec or Morelia exhibit much flatter time series. Changes in betweenness centrality for Valle de Mexico are very different to those for the other nodes. Evolution for degree and strength in Valle de Mexico seem to be correlated, but they differ from evolution of betweenness.

A relatively simple intuition is that municipalities with larger populations will have higher values of centrality measures regardless of time. We do see that all of the analysed centrality measures are correlated with population size. However, we observe that this correlation scales in the same way; Some of these metrics show a better lineal correlation with population on lineal scale while some other show a better lineal correlation on logarithmic scale (see Appendix for model fits).

A related question is whether variations in these centrality measure are also correlated with population size. We observe a negative correlation between coefficient of variation and population, showing that centrality variation is higher in smaller municipalities. Again, the best fit for this correlation is not necessarily found in the linear scale (see Appendix). In any case, it should be noted that while population size is clearly a relevant factor, the dispersion of the point cloud in the range of midsized municipalities indicate other factors that may be in play to fully explain centrality measures in these networks. 

Regarding formation of communities in the network, we observe that, in a very broad way, for each day in this analysis there is a ``giant'' community which includes about half or more of the nodes in the network and that this structure seems to preserve for all dates (see Fig. \ref{fig:alluvial_national}. Therefore, if there are changes in the community structure of inter-municipal mobility, these have to occur at more local levels. In this diagram lines (nodes in the network) are colored according to their state. For example, all light-blue lines starting at the bottom-left corner of the diagram correspond to Oaxaca municipalities, while green lines just above are Puebla municipalities. Notice here how all nodes in Puebla tend to stay inside the same community (the largest one), but municipalities in Oaxaca move to different communities, indicating a change in the structure of inter-municipal level mobility within this state. 

Alluvial plot on Fig. \ref{fig:alluvial_national} shows community structure and its time evolution for the hole country. We show in Fig. \ref{fig:alluvial_states} local versions of these plots for four different states. Campeche and Nuevo León are among top three states in the country for Gross Domestic Product per capita (the other being Mexico City, which is only one node in this analysis, therefore it lacks true \textit{intra-state} community structure, since all municipalities are collapsed in the same node); on the other hand, Oaxaca and-Chiapas are the two states with the lowest GDP per capita. 


\section*{Methods}

\subsection*{Mobile device location data} 
We used mobile device location data for the time period between 2020-01-01 and 2021-12-31 within Mexican territory provided by Veraset, a company that aggregates anonymized mobile device location data. This source dataset is provided as a table in which each record (called a \textit{ping}) contains the position (latitude and longitude) of a given (anonymized) device for a given timestamp (with temporal resolution up to seconds). The set of all unique device ids for a given day is called the \textit{device panel}.

\subsection*{Intermunicipal travel network construction} 
We define an Intermunicipal Travel Network (IMTN) as a directed, weighted graph G(V,E), for a given day, such that: 

\begin{itemize}
    \item nodes represent localities (either municipalities or metropolitan zones, as defined by the national geographic agency; see Note 1) 
    \item links represent mobility from the source node to the target node, defined by observing at least one device that moved from node i to node j. 
    \item Link weights represent the total fraction of observed devices that moved from node i to node j, out of the total number of observed devices; this acts as a normalized measure of flow between nodes (see Note 2). 
\end{itemize}

\subsubsection*{Note 1: on the aggregation of metropolitan area nodes}
The political division of Mexico has municipalities as the smallest unit. Generally, a population center is contained within a municipality; however, there are large urban areas in which a single population center extends through many different municipalities, such that the movement between municipal boundaries is capturing the urban mobility and not travel between different locations. The National Geography and Statistics Institute (INEGI) defines 74 metropolitan areas in Mexico, based on measurements from 2015 \url{https://www.inegi.org.mx/contenido/productos/prod_serv/contenidos/espanol/bvinegi/productos/nueva_estruc/702825006792.pdf}.
Since intracity mobility is beyond the scope of this manuscript (and would greatly skew the mobility metrics, as the volume of intra-city mobility is way larger than that of true travels between different locations), we decided to aggregate the municipalities that form these metropolitan areas into single nodes in the network.

\subsubsection*{Note 2: on the interpretation of edge weights in the network}
We defined edge weights in the network as follows: 


\begin{equation}
    W_{ij} = \frac{|D_{ij}|}{|D|}
\end{equation}

\noindent where $|D_{ij}|$ is the number of devices that were observed to move from i to j (that is, were observed in i and their next immediate ping was in j) and $|D|$ is the total number of devices in the day's device panel. In this way, the weight represents a normalized measure of flow between regions; we may observe that in limit cases, the weight will be zero when there is no movement observed from one municipality to the other, and the weight would be 1 if all observed devices within the country travelled from region i to region j (which would be a virtually impossible scenario). An advantage of using this approach is that it controls variability in the number of observed devices each day, allowing for comparisons between days. 

\subsubsection*{Network construction algorithm}
The following pseudocode shows the the algorithm used for network construction, starting with the daily mobile device location data table: 

\begin{algorithm}
	\caption{Intermunicipal Travel Network Construction }
	\begin{algorithmic}[1]
	
	    \State \textbf{inputs}
	    
	    \State \textit{pings} = data.frame(device, timestamp, lat, long) 
        \State \textit{panel} = {unique device set}
        \State \textit{mun}   = municipality and metropolitan zone polygon 
        \\
        \State \textbf{intermediate objects}
        \State \textit{transitions} =  data.frame(from, to)
        \\
        \For {ping \textit{in} pings}
        \State pings.mun = locate latitude, longitude \textit{in} mun 
        \EndFor
        \\
        \For {k \textit{in} panel}
        \State k.pings = pings[k]
		\State sort k.pings by timestamp
		\If{k.pings[t].mun != k.pings[t+1].mun}
		\State transitions.append(from = k.pings[t].mun, to = k.pings[t+1].mun) 
		\EndIf
		\EndFor
		\\
		\State \textit{imtn} = 
		\For {i,j \textit{in} transitions[from,to]}
		\State imtn[i,j].weight = $\frac{count(imtn[i,j])}{size(panel)}/$
        \EndFor
        \\
        \State \textbf{RETURN} \textit{imtn}

	\end{algorithmic} 
\end{algorithm} 



\subsection*{Network analysis}

Networks were analyzed using the igraph library, version 1.2.7 \cite{igraph}, for the R programming language, version 4.1.0.

\section*{Data availability}
The collection of 731 intermunicipal networks is publicly available on a OSF repository \url{http://dx.doi.org/10.17605/OSF.IO/42XQZ}. 
\begin{sloppypar}
\bibliographystyle{siam}
\bibliography{references}
\end{sloppypar}

\end{document}